\documentclass[pra,twocolumn,superscriptaddress]{revtex4}
\usepackage{amsmath}
\usepackage{amssymb}

\usepackage{amsmath,amssymb}
\usepackage[usenames]{color}
\usepackage{amssymb}
\usepackage{grffile}
\usepackage[pdftex]{graphicx}
\usepackage{amsmath, amstext, amssymb, amsfonts, amsxtra}
\usepackage{textcomp}
\usepackage{xspace} 

\newcommand{\sgz}{\hat{\sigma}^z}
\newcommand{\sgp}{\hat{\sigma}^+}
\newcommand{\sgm}{\hat{\sigma}^-}
\newcommand{\aop}{\hat{a}}
\newcommand{\adop}{\hat{a}^{\dagger}}
\newcommand{\bop}{\hat{b}}
\newcommand{\bdop}{\hat{b}^{\dagger}}

\newcommand{\Hop}{\hat{H}}
\newcommand{\Hdis}{\hat{H}^{{\rm dis}}}
\newcommand{\dw}{\delta}
\newcommand{\Dop}{\mathcal{D}}
\newcommand{\Ddis}{\mathcal{D}^{{\rm dis}}}
\newcommand{\rhoop}{\hat{\rho}}
\newcommand{\im}{{\rm i}}
\newcommand{\fidelity}{\mathcal{F}}

\begin{document}

\title{Population transfer via a dissipative structural continuum}
\author{Wei Huang}
\affiliation{Guangxi Key Laboratory of Optoelectronic Information Processing, Guilin University of Electronic Technology, Guilin 541004, China} 


\author{Shan Yin }
\email{syin@guet.edu.cn}
\affiliation{Guangxi Key Laboratory of Optoelectronic Information Processing, Guilin University of Electronic Technology, Guilin 541004, China}

\author{Baohua Zhu}
\affiliation{School of material science and engineering, Guilin University of Electronic Technology, Guilin 541004, China}

\author{Wentao Zhang}
\email{zhangwentao@guet.edu.cn}
\affiliation{Guangxi Key Laboratory of Optoelectronic Information Processing, Guilin University of Electronic Technology, Guilin 541004, China}

\author{Chu Guo}
\email{guochu604b@gmail.com}
\affiliation{Quantum Intelligence Lab, Supremacy Future Technologies, Guangzhou 511340, China}

\begin{abstract}   
We propose a model to study quantum population transfer via a structural continuum. The model is composed of two spins which are coupled to two bosonic modes separately by two control pulses, and the two bosonic modes are coupled to a common structural continuum. We show that efficient population transfer can be achieved between the two spins by using a multi-level stimulated Raman adiabatic passage (STIRAP) across the continuum, which we refer to as straddle STIRAP via continuum. We also consider the stability of this model against different control parameters and show that efficient population transfer can be achieved even in presence a moderate dissipation.
\end{abstract}

\date{\today}
\maketitle

\address{} 

\vspace{8mm}

\section{Introduction}
Complete population transfer serves transition population of quantum states from initial state to target state, which plays an important role in quantum physics. A lot of research efforts have been devoted to study complete population transfer in various situations. For instance, complete population transfer among quantum states of atoms and molecules is very active researching area in quantum optics and atom optics \cite{Kuklinski1989, Bergmann1998, Huang2017}. 
Furthermore, it is also a fundamental technique in quantum computation and quantum information processing, including superconducting qubits \cite{Falci2017,Chen2018,KumarParaoanu2016}, Bose-Einstein condensates \cite{Helm2018}, NV centers in diamond \cite{Chakraborty2017}, quantum dots and quantum wells in semiconductor \cite{Dory2016}. 
Another very important application of complete population transfer is to achieve power or intensity inversion in classical systems, which is widely used in waveguide couplers \cite{Huang2014}, wireless energy transfer \cite{Rangelov2011}, polarization optics \cite{Dimova2015} and electrons, surface plasmon polaritons in graphene system \cite{Huang2018sst, Huang2018carbon}. For a recent review one can refer to~\cite{Bergmann2019}.

A standard approach for population transfer is stimulated Raman adiabatic passage (STIRAP), which was originally proposed in three-level systems, two of which are coupled to an intermediate energy level by two spatially overlapping pulses in \textit{counter-intuitive} order. The remarkable dominance of STIRAP are that i) it is extremely robust against fluctuations of the control parameters of the laser pulses and ii) the intermediate energy level is not populated which makes the scheme robust against the decay \cite{Vitanov2001,Vitanov2017}. 

Various generalizations have made to apply STIRAP technique to special situations. STIRAP via multi-intermediate levels or continuum (multi-level STIRAP, also called straddle-STIRAP \cite{Vitanov1998}) has been considered in atomic system \cite{Peters2007, Vitanov1999, Rangelov2007} and waveguide couplers system \cite{Dreisow2009, Longhi2008}. STIRAP into continuum, where the third energy level is replace by continuous energy levels, has also been considered \cite{RangelovArimondo2010}.

In this paper, we propose a model to study population transfer via a continuum. The model contains two spins which are coupled to two bosonic modes separately by two controled laser pulses, while the two bosonic modes are indirectly coupled via a structured continuum. Compared to previous literatures, our model differs in that: i) the two energy levels are replaced by two spins, as a result, the population transfer becomes state transfer between the two spins; ii) instead of directly coupling the two energy levels with the continuum, in our approach the laser pulses directly couples the two spins with two bosonic modes, which could be single-mode cavities or phonons, and then the two bosonic modes are coupled to a continuum with constant coupling strengths; iii) dissipative continuum has been considered. This model has potential applications in chemical physics \cite{Deng2016} and quantum information \cite{Contreras2008}. In addition, this model allows us to study state transfer between two qubits via a dissipative environment, which could play an important role in quantum computation and quantum information processing. We demonstrate that straddle-STIRAP can be utilized to perform efficient population transfer in our model (see Fig.~\ref{fig:fig2} and Fig.~\ref{fig:fig3}). And we show the robustness of our approach with respect to parameters of controlling laser pulses (see Fig.~\ref{fig:fig4}) and dissipation rate (see Fig.~\ref{fig:fig5}).

Our paper is organized as follows. In Sec.\ref{sec:model}, we introduce our model and the equation of motion for the straddle-STIRAP via a continuum. In Sec.\ref{sec:result}, we numerically solve the quantum master equation for our model, and show the effectiveness of the popular transfer against changing the parameters of the model. We conclude in Sec.\ref{sec:summary}.

\section{Model}\label{sec:model}
\begin{figure}[hbtp]
\centering
\includegraphics[width=0.45\textwidth]{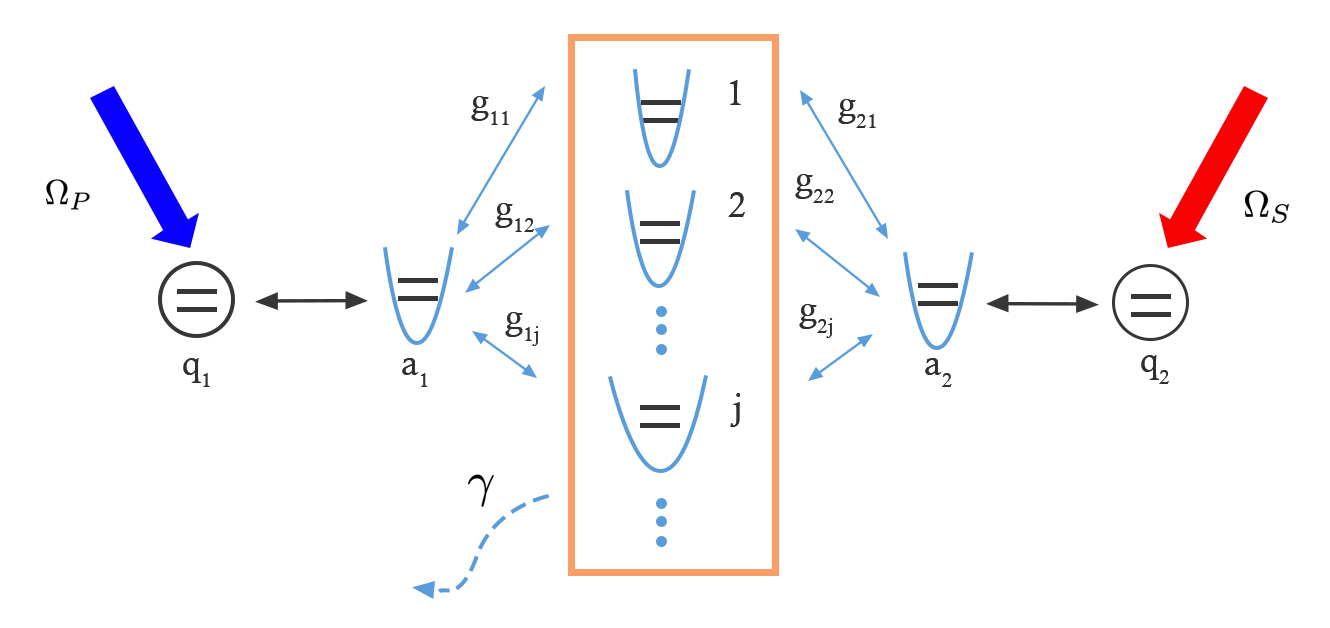}
\caption{Population transfer between two qubits $q_1$ and $q_2$. The two qubits are coupled to two bosonic modes $\aop_1$ and $\aop_2$ respectively by two controled laser pulses $\Omega_P(t)$, $\Omega_S(t)$. The bosonic modes $\aop_1$ and $\aop_2$ are indirectly coupled through a bosonic structural continuum, with a particle loss rate $\gamma$.} \label{fig:fig1}
\end{figure}

Our model consists of two spins which are coupled to two bosonic modes by two controled laser pulses. The two bosonic modes are both coupled to a bosonic continuum with phenomenological spectrum functions. The bosonic continuum is initially in the vacuum state and is subjected to a particle loss rate of $\gamma$. The Hamiltonian of the whole system can be written as
\begin{align} \label{eq:hamiltonian}
\Hop(t) =& \frac{\omega_{q,1}}{2}  \sgz_1 + \frac{\omega_{q,2}}{2}  \sgz_2 + \omega_{a,1} \adop_1\aop_1 + \omega_{a,2}\adop_2\aop_2 + \nonumber \\  
& \Omega_P(t) (\adop_1 \sgm_{1} + \aop_1 \sgp_{1}) + \Omega_S(t) (\adop_2 \sgm_{2} + \aop_2 \sgp_{2}) + \nonumber \\
&\int_{\omega} d\omega \omega \bdop(\omega) \bop(\omega) + \notag \\
& \int_{\omega} d\omega \sqrt{J_{1}(\omega)} \left(\aop_1 \bdop(\omega) + \adop_1 \bop(\omega)\right)  + \notag \\
& \int_{\omega} d\omega \sqrt{J_{2}(\omega)} \left(\aop_2 \bdop(\omega) + \adop_2 \bop(\omega)\right), 
\end{align}
where we have set $\hbar=1$. Here $\omega_{q,1}$ and $\omega_{q,2}$ are the energy differences of qubit 1 and qubit 2. $\omega_{a,1}$ and $\omega_{a,2}$ denote the oscillation frequencies of the two bosonic modes $\aop_1$ and $\aop_2$. $J_1(\omega)$ and $J_2(\omega)$ are the spectral densities for the coupling between the two modes $\aop_1$, $\aop_2$ and the bosonic continuum. We have used a linear density of states assumption for the continuum without loss of generality since the density of states can be absorbed into the spectral densities \cite{InesAlonso2016}. In this work we consider the phenomenological spectral densities which are defined as follows
\begin{align}
J_1(\omega) = g \omega^{\eta_1};
J_2(\omega) = g \omega^{\eta_2},
\end{align}
with a threshold $\omega_c$ such that $J_1(\omega) = J_2(\omega) = 0, \forall \omega > \omega_c$. The exponent $\eta < 1$, $\eta=1$ and $\eta>1$ correspond to the sub-ohmic, ohmic and super-ohmic couplings respectively. We also consider the situation where the bosonic continuum loses particles with a rate $\gamma$, which can be modeled by the Lindblad form of dissipation $\Dop$
\begin{align} \label{eq:dissipator}
\Dop(\rhoop) = \gamma\int_{\omega}d\omega \left[2\bop(\omega)\rhoop\bdop(\omega) - \{\bdop(\omega)\bop(\omega), \rhoop\} \right].
\end{align}
The dynamics of the system is thus described by the following quantum master equation
\begin{align} \label{eq:master}
\frac{d\rhoop(t)}{dt} = -\im\left[\Hop(t), \rhoop(t)\right] + \Dop(\rhoop(t)).
\end{align} 
Throughout this paper, we assume that $\omega_{q,1}=\omega_{q,2}=\omega_{a,1}=\omega_{a,2}=\Delta$. The initial state of the dynamical evolution is denoted as
\begin{align}
\rhoop_i = \rhoop(-\infty) = \vert \psi_i \rangle\langle \psi_i\vert,
\end{align}
with
\begin{align} \label{eq:psi_i}
\vert \psi_i \rangle = \vert 1^{q_1}, 0^{a_1}, \vec{\bold{0}}^b, 0^{a_2}, 0^{q_2} \rangle,
\end{align}
where we have use $1^{q_i}, i=1,2$ to denote the spin up state for the two spins $q_1$ and $q_2$, $0^{a_i}, i=1,2$  to denote the vacuum state for the two bosonic modes $\aop_1$ and $\aop_2$, and $\vec{\bold{0}}^b$ to denote the vacuum state for the bosonic continuum $\bop(\omega)$. The final state after the evolution is denoted as $\rhoop_f = \rhoop(\infty)$, while the targeting final state is written as 
\begin{align}
\bar{\rhoop}_f = \vert \psi_f \rangle\langle \psi_f \vert,
\end{align}
with 
\begin{align}
\vert \psi_f \rangle = \vert 0^{q_1}, 0^{a_1}, \vec{\bold{0}}^b, 0^{a_2}, 1^{q_2} \rangle.
\end{align}
We define $\fidelity_1(t)$ to be the fidelity between $\rhoop(t)$ and $\rhoop_i$
\begin{align}
\fidelity_1(t) = \langle \psi_i \vert \rhoop(t) \vert \psi_i \rangle,
\end{align}
which is the population of the density operator on on first spin $q_1$. We define $\fidelity_2(t)$ to be the fidelity between $\rhoop(t)$ and $\bar{\rhoop}_f$
\begin{align}
\fidelity_2(t) = \langle \psi_f \vert \rhoop(t) \vert \psi_f \rangle,
\end{align}
which is the population of the density operator on the second $q_2$. We denote $\fidelity = \fidelity_2(\infty)$. $\fidelity=1$ corresponds to complete population transfer, while $\fidelity<1$ corresponds to partial population transfer.

\section{Results} \label{sec:result}

\begin{figure} [htbp]
\centering
\includegraphics[width=0.5\textwidth]{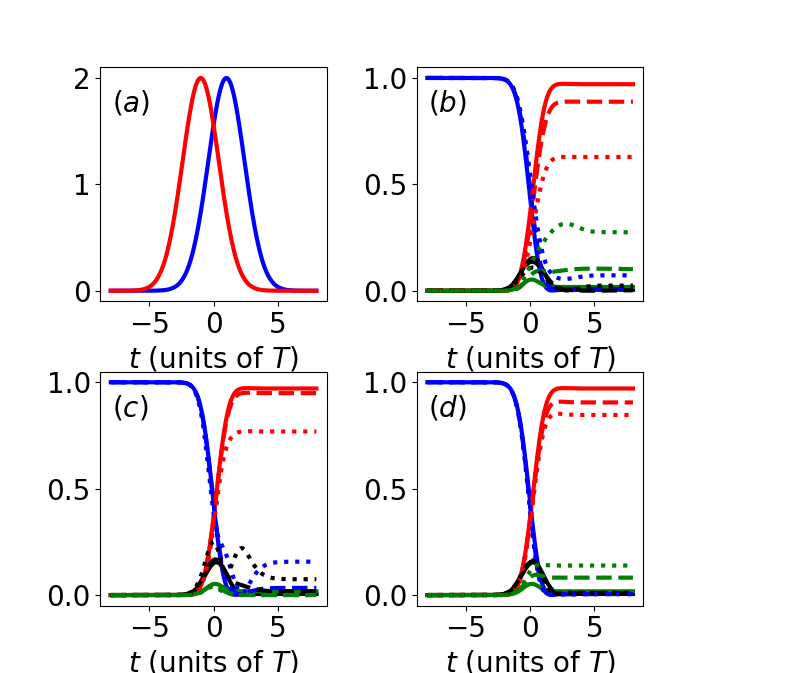}
\caption{(a) $\Omega_P(t)$ and $\Omega_S(t)$ as functions of time $t$. In (b), (c), (d), the evolution of $\fidelity_1(t)$ (blue lines) and $\fidelity_2(t)$ (red lines), as well as the population left in the two bosonic modes (black lines) and in the continuum (green lines) are plotted as a function of time. (b) We fix $\gamma=0$, $\Delta=0$, $\eta_1=1.5$. The solid, dashed and dotted lines correspond to $\eta_2=1.5, 1, 0.5$ respectively. (c) We fix $\gamma=0$, $\eta_1=\eta_2=1.5$. The solid, dashed and dotted lines correspond to $\Delta=0,5,10$ respectively. (d) We fix $\Delta=0$, $\eta_1=\eta_2=1.5$. The solid, dashed and dotted lines correspond to $\gamma=0, 0.5, 1.5$ respectively. The other parameters used are $g=10$, $\Omega=2$, $\omega_c=2$, $T=2$.} \label{fig:fig2}
\end{figure}

We numerically study the quantum master equation of Eq.(\ref{eq:master}). To numerically treat the bosonic continuum, we discretize it linearly with a discretization step size $\dw$, following~\cite{InesWolf2015}. The continuum becomes a discrete set of harmonic oscillators
\begin{align}
\int_{\omega} d\omega \omega \bdop(\omega) \bop(\omega) \rightarrow \sum_{j=1}^{N} \omega_j \bdop_j \bop_j, 
\end{align}
where $N=\omega_c/\dw$, $\omega_j = j\dw$, $\bop_j = \bop(j\dw)$ and $\bdop_j=\bdop(j\dw)$. The coupling between the bosonic modes and the continuum becomes 
\begin{align}
\int_{\omega} d\omega \sqrt{J_{1}(\omega)} \left(\aop_1 \bdop(\omega) + \adop_1 \bop(\omega)\right) \rightarrow \sum_{j=1}^N g_{1, j} \left(\aop_1 \bdop_j + \adop_1 \bop_j\right) \\
\int_{\omega} d\omega \sqrt{J_{2}(\omega)} \left(\aop_2 \bdop(\omega) + \adop_2 \bop(\omega)\right) \rightarrow \sum_{j=1}^N g_{2, j} \left(\aop_2 \bdop_j + \adop_2 \bop_j\right),
\end{align}
where the discretized coupling $g_{1,j}=\sqrt{J_1(j\dw)\dw}$, $g_{2,j}=\sqrt{J_2(j\dw)\dw}$. Combining the above equations, the discretized Hamiltonian is 
\begin{align}\label{eq:discreteH}
\Hdis(t) =& \frac{\Delta}{2}  \sgz_1 + \frac{\Delta}{2}  \sgz_2 + \Delta \adop_1\aop_1 + \Delta\adop_2\aop_2 + \nonumber \\  
& \Omega_P(t) (\adop_1 \sgm_{1} + \aop_1 \sgp_{1}) + \Omega_S(t) (\adop_2 \sgm_{2} + \aop_2 \sgp_{2}) + \nonumber \\
&\sum_{j=1}^{N} \omega_j \bdop_j \bop_j + \sum_{j=1}^N g_{1, j} \left(\aop_1 \bdop_j + \adop_1 \bop_j\right)  + \notag \\
& \sum_{j=1}^N g_{2, j} \left(\aop_2 \bdop_j + \adop_2 \bop_j\right). 
\end{align}
In the limit $N\rightarrow \infty$, $\Hdis(t)$ is equivalent to $\Hop(t)$~\cite{Bulla2008,InesWolf2015}.
The discretized dissipator can be simply written as
\begin{align}\label{eq:discreteD}
\Ddis(\rhoop(t)) = \gamma\sum_{j=1}^N \left[2\bop_j\rhoop\bdop_j - \{\bdop_j\bop_j, \rhoop\} \right].
\end{align}
We note that $\fidelity_1(t)$ and $\fidelity_2(t)$ should be independent of $\dw$ as long as $\dw$ is small enough. When $\gamma=0$, we directly solve the unitary dynamics with the time dependent Hamiltonian as in Eq.(\ref{eq:discreteH}). In case $\gamma >0$, we solve the quantum master equation in Eq.(\ref{eq:master}) with the discretized Hamiltonian as in Eq.(\ref{eq:discreteH}) and the discretized dissipator as in Eq.(\ref{eq:discreteD}). Although our model contains a large number of modes due to the continuum, it can be efficient solved by taking into account the fact that the model only contains at most $1$ excitation as can be seen from Eq.(\ref{eq:psi_i}), thus we only need to consider the vacuum sector together with the single excitation sector.

\begin{figure} [htbp]
\centering
\includegraphics[width=0.45\textwidth]{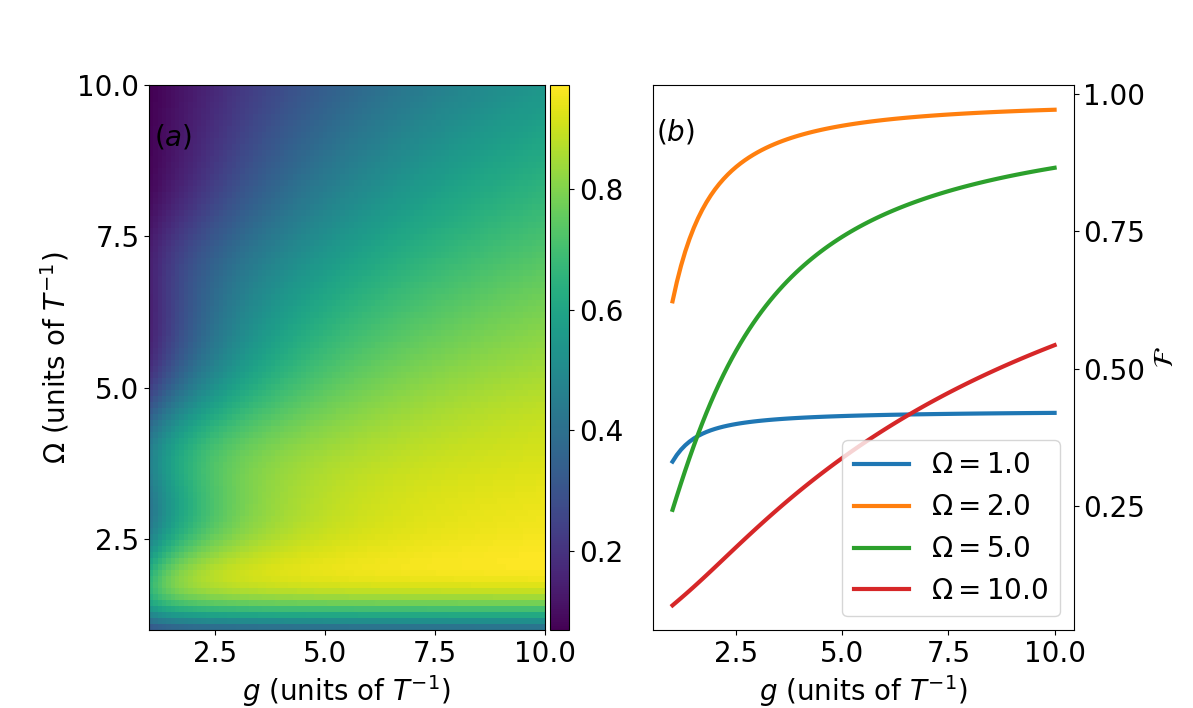}
\caption{(a) $\fidelity$ as a function of $\Omega$ and $g$. (b) Horizontal cuts of (a) for $\Omega=1, 2, 5, 10$ respectively. Other parameters used are $\Delta=0$, $\eta_1=\eta_2=1.5$, $\gamma=0$, $T=2$, $\tau=1$, $\omega_c=2$.} \label{fig:fig3}
\end{figure}

We consider that the two couplings of laser pulses ($\Omega_P$ and $\Omega_S$) have Gaussian shapes as follows
\begin{eqnarray}
&\Omega_P(t) = \Omega \exp\left(\dfrac{-\left(t-\tau /2\right)^2}{T^2}\right), \notag \\
&\Omega_S(t) = \Omega \exp\left(\dfrac{-\left(t+\tau /2\right)^2}{T^2}\right);
\end{eqnarray}
where the $T$ is the totally time for the control process, and $\Omega$ is the maximum strength of the coupling, $\tau$ is the time delay between two pulses. $\Omega_P(t)$ and $\Omega_S(t)$ are shown in Fig.~\ref{fig:fig2}(a). 

\begin{figure} [htbp]
\centering
\includegraphics[width=0.5\textwidth]{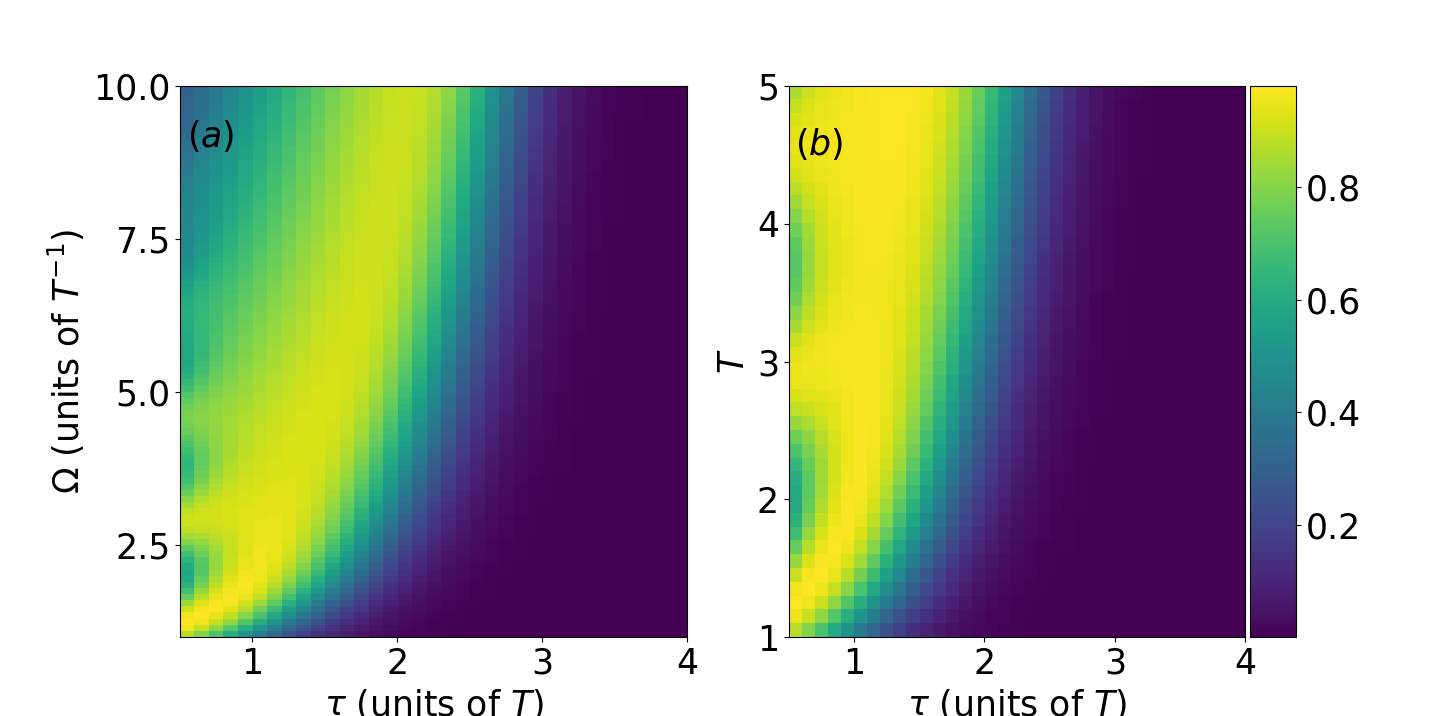}
\caption{(a) $\fidelity$ as a function of $\Omega$ and $\tau$, with maximum coupling strengths of laser pulses $\Omega$ from 1 $T^{-1}$ to 10 $T^{-1}$ and $\tau$ from 0.5 $T$ to 4 $T$, at fixed time $2T$. (b) $\fidelity$ as a function of $\tau$ and $T$, with $\tau$ from 0.5 $T$ to 4 $T$ and totally controlling time from 1 $T$ to 5 $T$, at fixed $\Omega = 2 T^{-1}$. } \label{fig:fig4}
\end{figure}

In Fig.~\ref{fig:fig2}(b), we consider the effect of asymmetric couplings between the two modes $\aop_1$, $\aop_2$ and the continuum, namely, $J_1(\omega) \neq J_2(\omega)$. We fix $\eta_1=1.5$, and tune $\eta_2$ to be $1.5, 1, 0.5$. We can see that $\fidelity$ is the largest when $\eta_1=\eta_2$, and decrease substantially when $\eta_2=0.5$, where $J_1(\omega)$ is super-ohmic while $J_2(\omega)$ is sub-ohmic, with a large portion of the population left in the continuum. It is shown in \cite{Vitanov1999} that when $J_1(\omega)$ and $J_2(\omega)$ are proportional to each other, complete population transfer could be achieved. Here we show numerically that when the couplings are asymmetric, the efficiency of population transfer could be greatly reduced. In Fig.\ref{fig:fig2}(c), we plot the evolution of the population of the two spins with $t$ against different values of $\Delta$, namely $\Delta=0, 5, 10$. We can see that $\fidelity$ greatly decreases when $\Delta$ is much larger than $\omega_c$, and a large portion of the population is left in the bosonic modes instead of the continuum in comparison with the previous case. This is because the spins are off resonant with the continuum and the population transfer is much harder (population transfer is still possible when $\Delta > \omega_c$ because of the strong coupling between the bosonic modes and the continuum). In Fig.\ref{fig:fig2}(d), we show $\fidelity$ against different particle loss rate, namely $\gamma = 0$ (solid line), $\gamma = 0.5$ (dashed line) and $\gamma =1$ (dotted line). As expected, population transfer becomes less efficient as $\gamma$ increases.

In Fig.~\ref{fig:fig3}, we study the effect of the competition between the two coupling strengths $\Omega$ and $g$ on the efficiency of the population transfer. In Fig.~\ref{fig:fig3}(a), we plot $\fidelity$ as a function of the $\Omega$ and $g$. We can see that when $\Omega \ll g$, efficient population transfer could be achieved, namely $\fidelity \approx 1$. To see this more clearly, in Fig.~\ref{fig:fig3}(b), we plot horizontal cuts of Fig.~\ref{fig:fig3}(a) at different values of $\Omega$, namely $\Omega=1,2,5,10$. 

\begin{figure} [htbp]
\centering
\includegraphics[width=0.45\textwidth]{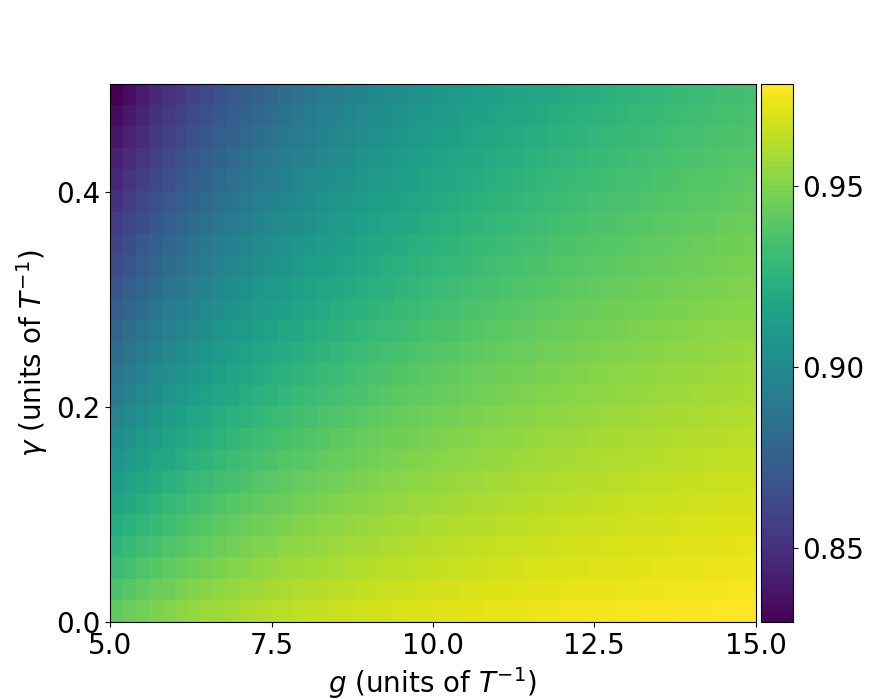}
\caption{The $\fidelity$ as a function of $g$ (coupling strengths between two qubits and spin bath) and dissipation loss $\gamma$.} \label{fig:fig5}
\end{figure}

Now we consider the robustness of our straddle STIRAP against the control parameters $T$, $\Omega$ and $\tau$ of laser pulses $\Omega_P(t)$, $\Omega_S(t)$, which is shown in Fig.~\ref{fig:fig4}. In Fig.~\ref{fig:fig4}(a), we shown the dependency of $\fidelity$ as a function of $\Omega$ and $\tau$, where we can see that population transfer can still be achieved with high efficiency if the values of $\Omega$ and $\tau$ has small fluctuations. In Fig.~\ref{fig:fig4}(b), we can see that for fixed $\Omega=2$, and $\tau \approx 1$, population transfer is highly efficient for a very wide range of $T$. We also notice that for small values of $\tau$, namely $\tau \approx 0.5$, there are some oscillations for certain values of $\Omega$ and $T$. A possible reason for these oscillations is that when $\tau$ is small, the evolution is non-adiabatic, and for certain special values of $\Omega$ and $T$, some non-adiabatic shortcuts lead to similar results as the adiabatic evolution.

Finally, we study the effect of dissipation on the straddle STIRAP. We assume the bosonic continuum has a constant particle loss rate $\gamma$. In Fig.~\ref{fig:fig5}, we plot the dependency of $\fidelity$ as a function of the particle loss rate $\gamma$ and the coupling strength $g$ between the modes $\aop_1$, $\aop_2$ with the continuum. We can see that as long as the coupling strength $g$ is large enough $g\geq 10$, efficient population transfer can still be achieved with moderate dissipation $\gamma \leq 1$.

\section{conclusion} \label{sec:summary}

We propose a model to study population transfer where the intermediate states is a bosonic continuum. The model consists of two spins which are coupled to two bosonic modes with a dynamical coupling strength $\Omega_P(t)$ and $\Omega_S(t)$, and the two bosonic modes are indirectly coupled through a bosonic continuum. We show the effects on the efficiency of population transfer when tuning the the coupling strength between the bosonic modes with the continuum, as well as the various control parameters of the laser pulses. We also consider the case that when the continuum subject to a constant particle loss rate, and show that efficient population transfer can still be achieved with a moderate dissipation. We believe that this finding will be improve the high efficient transfer information in quantum information processing in future.

\section{Acknowledgement}
We thanks the useful discussions with Prof. Tim Byrnes (Shanghai NYU) and Prof. Jonathan P. Dowling (Louisiana State University).

This work is acknowledged for funding National Science and Technology Major Project (grant no. 2017ZX02101007-003); National Natural Science Foundation of China (grant no. 61565004; 6166500; 61965005); the Natural Science Foundation of Guangxi Province (Nos. 2017GXNSFBA198116 and 2018GXNSFAA281163); the Science and Technology Program of Guangxi Province (No. 2018AD19058). W.H. is acknowledged for funding from Guangxi oversea 100 talent project and W.Z. is acknowledged for funding from Guangxi distinguished expert project.


\end{document}